\documentclass[floatfix,aps,showpacs,twocolumn]{revtex4}
\usepackage{graphicx}
\begin{document}

\title{Logarithmic rate dependence in deforming granular materials}

\author{R.~R. Hartley}
\affiliation{Department of Physics \& Center for Nonlinear and Complex
Systems, Duke University, Durham NC, 27708-0305, USA}

\author{R.~P. Behringer}
\email{bob@phy.duke.edu}
\affiliation{Department of Physics \& Center for Nonlinear and Complex
Systems, Duke University, Durham NC, 27708-0305, USA}

\date{Oct. 8, 2002}

\begin{abstract}

Rate-independence for stresses within a granular material is a basic
tenet of many models for slow dense granular
flows\cite{wood90,nedderman92,jaeger96,herrmann97,behringer97}. By
contrast, logarithmic rate dependence of stresses is found in
solid-on-solid friction\cite{ovarlez01,nasuno98,losert00}, in
geological settings\cite{dieterich79,ruina83}, and
elsewhere\cite{berthoud99,baumberger99,carlson96,gnecco00}. In this
work, we show that logarithmic rate-dependence occurs in granular
materials for plastic (irreversible) deformations that occur during
shearing but not for elastic (reversible) deformations, such as those
that occur under moderate repetitive compression. Increasing the
shearing rate, $\Omega$, leads to an increase in the stress and the
stress fluctuations that at least qualitatively resemble what occurs
due to an increase in the density. Increases in $\Omega$ also lead to
qualitative changes in the distributions of stress build-up and
relaxation events. If shearing is stopped at $t=0$ , stress
relaxations occur with $\sigma(t)/\sigma(t=0) \simeq A \log(t/t_o)$ .
This collective relaxation of the stress network over logarithmically
long times provides a mechanism for rate-dependent strengthening.

\end{abstract}


\maketitle

Slow granular flows, the subject of this letter, are typically
described in the context of Mohr-Coulomb friction
models\cite{nedderman92} that resemble those used for describing
friction between two solid bodies\cite{heslot94}.  In the well-known
solid friction scenario\cite{persson98,meyer98}, an object on a
frictional surface will resist a force and remain at rest provided the
magnitude of the tangential force is less than the product of a static
friction coefficient and the normal force.  This picture was
translated into the granular context (for dense granular systems
characterized by networks of force chains\cite{howell99a}) by
Coulomb\cite{coulomb76} and more recent
authors\cite{wood90,nedderman92}, where the normal and tangential
forces are replaced by corresponding normal and shear stresses, and
the surface of interaction is replaced by a plane within the
material. For large enough tangential force (shear stress) relative to
the normal force (normal stress) sliding friction (failure in a
granular material) occurs.  In these pictures, sliding friction
(deformation following failure) is independent of the speed of sliding
(the shear rate).

In reality, experiments in diverse contexts have shown that solid
friction exhibits a logarithmic dependence on rate and that static
frictional contacts strengthen logarithmically with age.  These
experiments span a vast range of lengths, and include studies at the
atomic\cite{gnecco00},
lab\cite{nasuno98,losert00,berthoud99,baumberger99,carlson96}, and
geological scale\cite{dieterich79,ruina83}. Recent experiments by
Ovarlez \emph{et al.}\cite{ovarlez01} using granular materials sliding
against the interior wall of a piston showed clear rate dependence
that was associated with aging effects of individual solid-friction
contacts and with the force network.  Additionally, experiments by
Nasuno \emph{et al.}\cite{nasuno98,losert00} in which a solid surface
was pushed across a granular bed showed a slow strengthening (or
aging) with time of the (quasi-static) force.  In the present
experiments, which zoom in uniquely on the grains, we show that there
is a logarithmic rate-dependence in slowly sheared granular materials
that is associated with irreversible rearrangements of the grain
contacts within the material itself.  This is manifest as a
strengthening with rate, and it is unique to granular systems and
possibly other jammed systems\cite{liu98}; i.e., it does not depend
\emph{per se} on the logarithmic strengthening of frictional contacts. 

\begin{figure}
\includegraphics[width=\hsize]{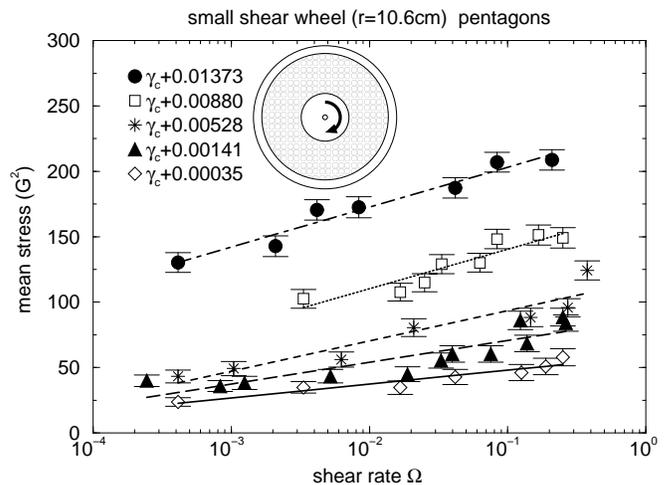} 
\caption{Mean stress vs. shearing rate, $\Omega$ for a range of
densities for shearing.  The lin-log scales emphasize the fact that
the data is consistent with a logarithmic variation of the stress with
$\Omega$.  Inset: sketch of Couette shear apparatus.}
\label{fig:mean} 
\end{figure}

The experiments described here were carried out with a 2D realisation
of a granular system.  The particles were made of a photoelastic
material and were either relatively thin disks or flat particles with
a pentagonal cross section.  By using a photoelastic material, we
could determine forces at the grain scale.  More detailed descriptions
of the experimental apparatus and methods\cite{heywood52} have been
given elsewhere for related
experiments\cite{howell99a,howell99b,geng01}, and here we provide only
essential information for two qualitatively different experiments. For
each experiment, the humidity varied by no more than $\pm 5\%$.  In the
first of these two experiments, the photoelastic particles were
sheared in an annular geometry (inset, Fig.~\ref{fig:mean}) that was
bounded on the inside by a rough wheel, the source of shearing, and on
the outside by an equally rough static ring.  The particles sat on a
horizontal, flat powder-lubricated Perspex sheet, and were contained
from above by a similar sheet. The whole was placed in a polariscope
consisting of a pair of right and left circular polarizers, a light
source and a video camera. Photoelastic images were obtained at rates
up to 15~Hz, captured by a framegrabber, and processed on-the-fly to
determine the forces on the particles within the field of view of the
camera ($\sim 100$ to $\sim 200$ particles).  Specifically, we
computed the force on each particle in each frame and summed over all
particles in a frame.  We stored this integrated force for each image
to produce a time series over very long sampling times, between
$2^{14}$ and $2^{16}$ points depending on the rate of shearing. To a
reasonable approximation ($\sim 10$\%), this quantity is proportional
to the pressure, and for simplicity, we will refer to it as the
stress.

Key control parameters are the rate of shearing, $\Omega = 2\pi/T$,
where $T$ is the period for one rotation of the shearing wheel, and
the density of the system, which we give in terms of the area
fraction, $\gamma$, occupied by the particles.  As shown
previously\cite{howell99a,howell99b}, there is a critical value of
$\gamma_c$, below which the shearing ceases, and we typically
reference $\gamma$ to this value.  ($\gamma_c$ depends on properties
such as the particle shape and possibly the geometry of the
container.)  It is also useful to give the rotation rate,
$f=1/T=\Omega/2\pi$. In these experiments, $0.03 \textrm{ mHz} \leq f
\leq 60 \textrm{ mHz}$.

\begin{figure}
\includegraphics[width=\hsize]{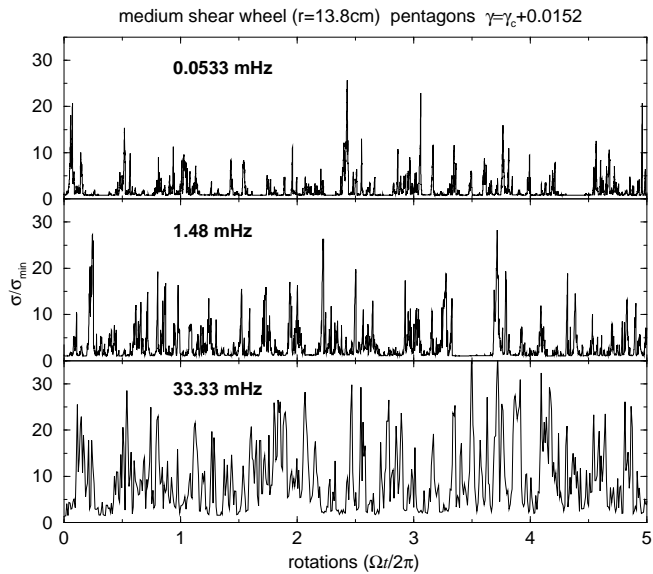} 
\caption{Time series for stress (in arbitrary units) for a range of
shearing rates $\Omega$ that span the experimental range.  These data
are for pentagonal particles, but data for disks are qualitatively
similar.}
\label{fig:time_series} 
\end{figure}

In Fig.~\ref{fig:time_series}, we show several examples of stress time
series for $\Omega$'s ranging over about three orders of magnitude in
the shearing rate.  The three images of this figure give an indication
of the effect of shearing rate: the time series for the slowest rate
is clearly of lower amplitude and more intermittent than that for the
highest rate.

The rate dependence is seen clearly in the mean (time-averaged) stress
vs. $\Omega$, Fig.~\ref{fig:mean}.  These data are consistent with a
logarithmic rate dependence for all $\gamma$'s. Similar effects have
been seen in solid-on-solid friction experiments but with a
significant difference: in the present experiments, there is a
logarithmic strengthening of the mean stress with rate, while other
experiments\cite{ovarlez01,losert00,berthoud99} show a logarithmic
strengthening with waiting time and/or humidity.  If we fit these data
to the form $\bar{\sigma} = A \log(\Omega/\Omega_o)$, then the
amplitude, $A$, is sensitive to the density, but the characteristic
frequency, $\Omega_0$, depends only weakly on $\gamma$, and
corresponds to a typical time of $2\pi /\Omega_o \approx 30 \textrm{
days}$.  However, there is no compelling reason to believe that this
relation holds in the limit $\Omega \to 0$.  A more realistic
description may be $\bar{\sigma} = A \log[(\Omega +
\Omega_1)/\Omega_o]$, where $\Omega_1>\Omega_o$.

\begin{figure}
\includegraphics[width=\hsize]{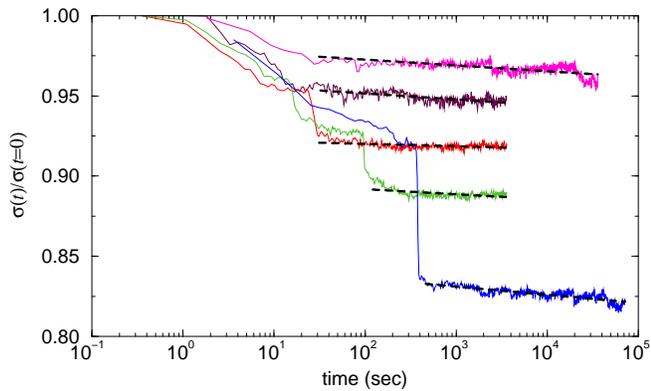} 
\caption{Relaxation of stress vs. time of  static
configurations of grains, previously undergoing plastic deformation
due to shear. The dashed lines are a fit to $\sigma(t)/\sigma(t=0)
\simeq A \log(t/t_o)$}
\label{fig:decay} 
\end{figure}

If a sample is sheared under steady state conditions and the shearing
wheel is abruptly stopped, the stress relaxes over very long time
scales.  In Fig.~\ref{fig:decay}, we show the relaxation of stress
vs. time on a semi-log plot.  Note that there are two time scales: a
quick initial decay lasting $\approx 20\textrm{ s}$ and a much longer
process such that the stress network may still be relaxing after
20~hrs.  Fig.~\ref{fig:decay} shows fits to the stress decrease with
time, of the form $\sigma (t=0)/\sigma (t)\simeq A \log (t/t_{0})$.
Although the relaxation is not completely uniform in time, a
logarithmic functional form is not unreasonable for many cases (after
an initial rapid relaxation).  This was not always the case, and an
interesting example of a major relaxation event is seen in one of
these runs about 350~s after the stopping of shearing; thereafter,
continued slow relaxation occurred.  A key insight is that, while the
micro-contacts between particles may be strengthening in
time\cite{losert00,berthoud99,heslot94}, the stress network as a whole
is relaxing over very long time scales.

\begin{figure}
\includegraphics[width=\hsize]{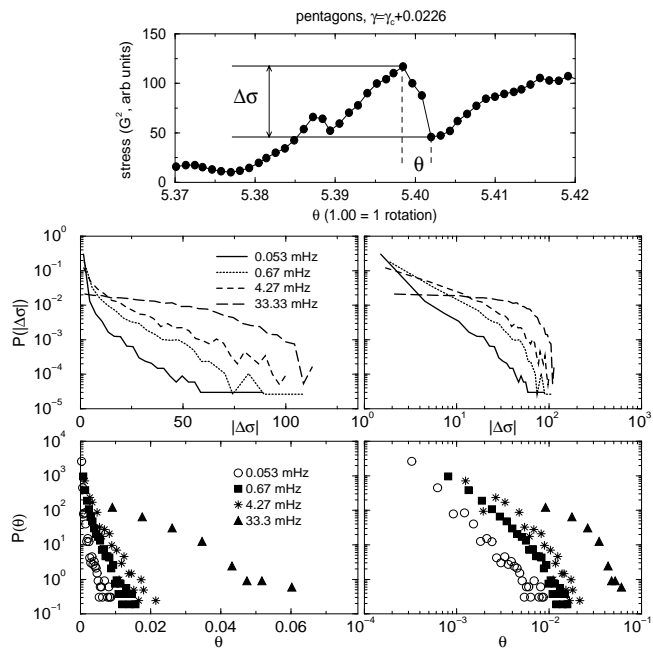} 
\caption{Top: a short segment of a time series that identifies a
single avalanche, $\Delta \sigma$ and $\theta$. Bottom: distributions
of stress avalanches for a dense system of pentagons, $\gamma =
\gamma_c+0.0226$, for several rates on a log-lin graph (left) and a
log-log graph (right).}
\label{fig:aval_dists}
\end{figure}

We gain additional insight into the effect of rate changes by
determining the distribution of build-up events (monotonic increases
in stress) and release events ("avalanches" -- monotonic decreases in
stress). Here, there are two aspects that are of interest: the size of
the build ups and avalanches, $\Delta \sigma$, and the angle,
$\theta$, through which the shearing wheel turns during the course of
a build-up or an avalanche. In the top part of
Fig.~\ref{fig:aval_dists}, we show an example of an "avalanche" event,
and in the bottom, we show distributions for $\Delta \sigma$ and
$\theta$ for avalanches.  Note that for near-critical $\gamma$'s and
for the slowest $\Omega$'s, the avalanche distributions are consistent
with power laws.  At higher $\gamma$ and $\Omega$ the distributions
for both build-up (not shown due to space constraints) and avalanche
events are roughly exponentials.  Interestingly, both increases in
$\Omega$ and in $\gamma$ tend to have the same qualitative effect on
the statistics of avalanches and build-ups.

\begin{figure}
\includegraphics[width=\hsize]{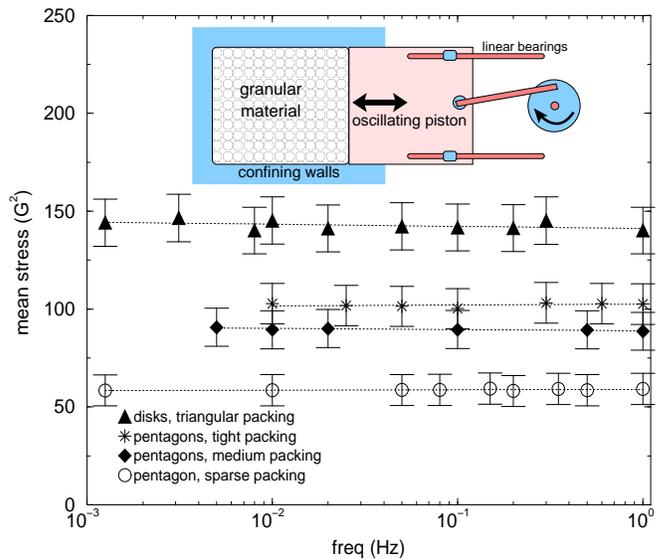} 
\caption{Mean stress vs. oscillation rate, $\Omega$ for several densities
for oscillatory compression.  Data for both disks and pentagons is
shown.  The lin-log scales emphasize the fact that there is no
variation of the stress with $\Omega$ within experimental
uncertainties.  Inset: sketch of piston apparatus.}
\label{fig:piston} 
\end{figure}

In order to further explore the origin of this rate dependence, we
have carried out a second set of experiments, using the same
photoelastic techniques, in which the particles were confined to a box
with three fixed sides.  The sample was then compressed and released
repeatedly on the fourth side by means of an oscillating piston, as in
the sketch of the inset of Fig.~\ref{fig:piston}.  The driving
velocity of the piston was chosen to match the corresponding
velocities of the shear wheel in the first experiment.  The key
difference between the two sets of experiments was that in the first
set there was continuous plastic deformation while in the latter set
there was none; i.e., the particles maintained their relative
position.  As shown in the body of the figure, which gives
$\bar{\sigma} $ vs. $\Omega$, there was no rate dependence of this
process, within experimental error.

To conclude, we have demonstrated that sheared granular materials
exhibit logarithmic rate dependence that is tied in an essential way
to particle rearrangements (plastic deformation).  The nature of this
rate dependence is further illuminated by the slow relaxation that
occurs when shearing is halted.  The latter experiments in particular
suggest that there are slow collective rearrangements of the particles
that can occur over large time scales.  This slow relaxation of
stresses over long times is consistent with increases in with rate:
force chains generated by shearing cannot completely relax on the time
scales over which new chains are formed, an effect that is exacerbated
by increasing $\Omega$. It seems likely that this effect is intimately
related to the long time scales reported for compaction in granular
materials\cite{nowak98}.  For compaction, collective rearrangements
become progressively more difficult over time because each new
rearrangement requires the involvement of larger collections of
particles.  Similarly, in the present experiments, relaxation events
involving collections of particles become less probable over time due
to geometric effects and to the fact that the available elastic energy
is gradually reduced with each rearrangement. An interesting question
is whether similar properties are seen in other jammed
systems\cite{liu98} such as colloids or foams.

\begin{acknowledgments}
Acknowlegements: This work was supported by NSF Division of Materials
Research, NSF Division of Materials Science \& by NASA. We appreciate
helpful conversations with Prof. Sue Coppersmith.
\end{acknowledgments}

\end{document}